\documentclass[twocolumn,english,reprint,amsmath,amssymb,onecolumn,nofootinbib,11pt]{revtex4-2}
\usepackage[T1]{fontenc}
\usepackage[latin9]{inputenc}
\makeatletter
\usepackage{float}
\usepackage{physics}
\usepackage{amsfonts}
\usepackage{latexsym}
\usepackage{epsfig}
\usepackage{graphicx}
\usepackage{tikz}
\definecolor{greatblue}{RGB}{40,120,181}
\definecolor{greatred}{RGB}{200,36,35}
\usepackage[colorlinks,linkcolor=greatblue,anchorcolor=blue,citecolor=greatred]{hyperref}
\usepackage{dcolumn}
\usepackage{bm}
\makeatother
\usepackage[english]{babel}

\begin{document}

\title{Finite Modular Hamiltonians and Their Singular Limits}

\author{Panyinyan Li}
\email{lipanyinyan@stu.scu.edu.cn}
\affiliation{College of Physics, Sichuan University, Chengdu, 610065, China}

\author{Jun Tao}
\email{taojun@scu.edu.cn}
\affiliation{College of Physics, Sichuan University, Chengdu, 610065, China}

\author{Haitang Yang}
\email{hyanga@scu.edu.cn}
\affiliation{College of Physics, Sichuan University, Chengdu, 610065, China}

\begin{abstract}
Using the open solid torus as a finite parent, we revisit modular flow for spherical regions in conformal field theory. Its nonsingular angular evolution gives   finite modular generators in a very simple form, while an exact conformal map to a cavity frame recovers the Rindler, CHM, and stationary-source ball constructions  as singular limits. The same finite parent also provides a regular completion of the center obstruction in the complexified ball flow. Charged, Dirac, and inhomogeneous scalar examples are given.
\end{abstract}

\maketitle
\tableofcontents

\section{Introduction}

In Ref.~\cite{Jiang:2026all}, the entanglement entropy of a spherical region in a CFT was calculated in arbitrary spacetime dimension by studying the CFT on an open solid torus (OST) as depicted in Fig. \ref{fig:ost}. The simplicity of that calculation came mainly from the geometry. The OST is open. The polar fixed locus does not belong to the physical manifold. Its replica has exactly the same shape: only the angular period changes from $2\pi$ to $2\pi n$. No conical fixed point has to be inserted into the physical CFT region. After a global conformal transformation the finite OST becomes a spherical cavity  as depicted  in Fig. \ref{fig:cavity}, and the conventional planar configuration is recovered as a   singular limit $\epsilon\to 0$. This order of operations was the essential reason that the intermediate quantities remained finite and simple.

The same structure appeared again in a different problem. In Ref.~\cite{Jiang:2026exact}, the Weyl-frame two-point function of a CFT on the OST was shown to be exactly paired with a finite geodesic in the associated AdS geometry. The usual boundary-anchored asymptotic geodesic relation is obtained as a singular limit of the finite configuration.
Together with the finite entanglement construction, this led us to the perspective developed further in Ref.~\cite{Yang:2026kin}: conventional singular or asymptotic objects should, whenever possible, be regarded as singular limits of finite parents. The practical rule is simple: rather than regularizing a singular object from the beginning, first look for a finite parent from which it descends.

This viewpoint is useful for more than removing divergences. A finite parent can reorganize a problem so that structures which look complicated in the singular representation become simple and elementary before the limit is taken. In suitable cases, calculations that are inaccessible in the singular variables become workable finite calculations. It can also give a well-defined meaning to objects that are awkward or pathological in the conventional singular variables. The point is not that every singularity of a  system disappears. Rather, the finite parent provides a regular object before the singular representation is introduced, and therefore separates genuine physics from singularities created by a degeneration or by an incomplete coordinate patch.

The purpose of the present paper is to apply this idea to modular Hamiltonians and modular flow. Modular Hamiltonians are in general nonlocal and difficult to construct. Important exceptions include the Bisognano--Wichmann boost generator for a vacuum wedge \cite{Bisognano:1975ih,Bisognano:1976za}, conformal modular actions for double cones \cite{Hislop:1981uh,Brunetti:1993dx}, and the spherical CFT construction of Casini, Huerta and Myers (CHM) \cite{Casini:2011kv}. Euclidean-source states and deformations have also been studied systematically \cite{Lashkari:2016vgj,Faulkner:2016mzt,Balakrishnan:2020lbp,Arias:2020mhe}, together with more general questions of local and geometric modular flow \cite{Arias:2016nip,Casini:2017roe,Sorce:2024kgg,Caminiti:2025gmf}. 
In this work, we show that the OST gives them a common finite origin.

The finite-parent picture is especially transparent here. On the OST the angular direction is globally defined and has no fixed point inside the physical domain. Cutting the angular path integral therefore gives an ordinary transfer matrix. For a local preparation that is invariant under the angular rotation, the modular generator is simply the corresponding angular Hamiltonian. The entire construction is finite. A conformal transformation then transports the result to a finite spherical cavity configuration  in Fig. \ref{fig:cavity}. Only at the end is the cavity collapsed to the conventional ball.

This perspective also clarifies a specific difficulty encountered in the conventional ball description. Arias, Botta-Cantcheff, Martinez and Zarate observed that the analytical extension of the modular flow in the ball is ill-defined at the center $x^i=0$. They thus introduced the punctured region $V_0=V\setminus\{0\}$ in their complexified construction \cite{Arias:2020mhe}. In the finite OST representation no point has to be removed: the corresponding angular orbit is a smooth circle in the parent. The apparent obstruction appears only after the conformal map to the ball variables, where part of this smooth orbit is pushed to conformal infinity. Thus the finite parent gives a regular completion of the modular trajectory even when a single ball coordinate chart is not global. We will make this relation explicit below.

The same finite construction reproduces the other standard ingredients of CHM. In the singular limit the finite hyperbolic section becomes the complete $H^{d-1}$, angular evolution becomes ordinary time translation on $\mathbb R\times H^{d-1}$, the vacuum reduced density matrix becomes unitarily equivalent to a thermal density matrix at temperature $1/(2\pi R)$, the familiar local ball modular Hamiltonian is recovered, and the exact finite relation between the cavity scale and hyperbolic radius reduces to the usual UV/IR relation. 

We provide several examples to show how the same organization works beyond the vacuum. A constant $U(1)$ chemical potential gives a charged finite modular Hamiltonian and reduces in the singular limit to the standard charged spherical construction \cite{Belin:2013uta,Dowker:2015dpd}. A four-dimensional massless Dirac field provides an explicit spectral benchmark. Finally, a localized quadratic shell in the three-dimensional conformal scalar gives a spatially inhomogeneous stationary state whose modular Hamiltonian is immediate on the finite parent, while the remaining normalization problem reduces, in the complete-hyperbolic limit, to an ordinary scattering problem.

The paper is organized as follows. Section II reviews the OST as a finite parent and derives the angular transfer matrix. Section III maps the construction to a finite spherical cavity and shows how the CHM results arise in the singular limit. Section IV discusses the complexified-flow issue and stationary local sources, including the relation to Ref.~\cite{Arias:2020mhe}. Section V gives the conserved-charge and Dirac examples. Section VI considers the inhomogeneous scalar shell. Section VII summarizes the finite-parent interpretation. Technical details are collected in the appendices.

\section{Open-Solid-Torus as a Finite Parent}
\label{sec:ost}

\subsection{Open geometry and angular evolution}

\begin{figure}[t]
\centering
\includegraphics[width=0.3\textwidth]{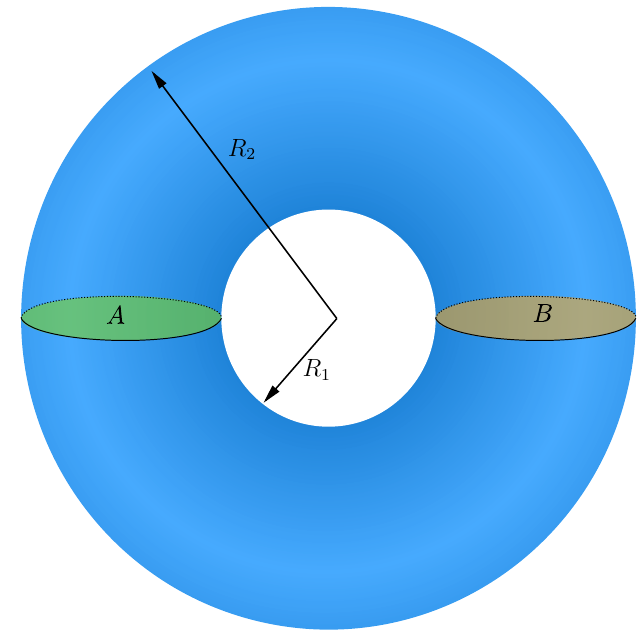}
\caption{The CFT lives on a flat open solid torus (OST). The angular circle is everywhere nondegenerate in the physical domain. The regions $A$ and $B$ are disjoint at finite separation. 
}
\label{fig:ost}
\end{figure}

Consider a $d$-dimensional Euclidean CFT in flat space,
\begin{equation}
 ds_E^2=dt_E^2+dy^2+d\bm{x}_\perp^{\,2},
 \qquad \bm{x}_\perp\in\mathbb R^{d-2}.
\label{eq:flatmetric}
\end{equation}
Introduce polar coordinates in the $(t_E,y)$ plane,
\begin{equation}
 t_E=\rho\sin\theta,\qquad
 y=\rho\cos\theta,\qquad
 \theta\sim\theta+2\pi,\qquad \rho>0.
\label{eq:polar}
\end{equation}
A reflection-symmetric open solid torus (OST) as depicted in Fig. \ref{fig:ost} is the domain
\begin{equation}
 \left(\rho-\frac{R_1+R_2}{2}\right)^2+|\bm{x}_\perp|^2
 <\left(\frac{R_2-R_1}{2}\right)^2,
 \qquad R_2>R_1>0.
\label{eq:ostdomain}
\end{equation}
Note the  inequality is strict and the OST has no boundary. The OST is flat.  Since $\rho=0$ lies outside the physical domain, the angular vector $\partial_\theta$ has no fixed point on the OST. 
This simple fact is useful both for replicas and for modular evolution. Replicating the OST keeps its shape and changes only the angular identification,
\begin{equation}
 \theta\sim\theta+2\pi n.
\label{eq:replicaperiod}
\end{equation}
There is no physical fixed locus at which a cone has to be inserted. This is the geometric origin of the simplification used in Ref.~\cite{Jiang:2026all}. In the present paper we use the same structure  to study the modular operator itself.

\subsection{Weyl frame and finite hyperbolic section}

The flat metric can be put into the Weyl frame by
\begin{equation}
 ds_E^2=\rho^2 ds_W^2,
 \qquad
 ds_W^2=d\theta^2+\frac{d\rho^2+d\bm{x}_\perp^{\,2}}{\rho^2}.
\label{eq:weylmetric}
\end{equation}
The OST therefore becomes
\begin{equation}
 {\cal M}_{\rm OST}=S^1_\theta\times Q_H,
\label{eq:product}
\end{equation}
where $Q_H$ is a finite geodesic ball in $H^{d-1}$. Let
\begin{equation}
 \rho_c=\sqrt{R_1R_2},
\end{equation}
and define the hyperbolic distance $u$ from $(\rho_c,\bm 0)$ by
\begin{equation}
 \cosh u=\frac{\rho^2+\rho_c^2+|\bm{x}_\perp|^2}{2\rho\rho_c}.
\label{eq:hyperdistance}
\end{equation}
As shown in Appendix~\ref{app:coordinates}, the OST leads to $0\le u<u_0$ with
\begin{equation}
 u_0=\frac12\log\frac{R_2}{R_1}.
\label{eq:u0}
\end{equation}
The Weyl-frame metric in this geodesic coordinate is
\begin{equation}
 ds_W^2=d\theta^2+du^2+\sinh^2u\,d\Omega_{d-2}^2,
\label{eq:globalH}
\end{equation}
with a finite hyperbolic volume
\begin{equation}
 V_H=\Omega_{d-2}\int_0^{u_0}du\,\sinh^{d-2}u.
\label{eq:VH}
\end{equation}
The coordinate derivation is given in Appendix~\ref{app:coordinates}.

\subsection{Cut and transfer matrix}

Cut the Euclidean path integral at a constant angular slice. If the theory is independent of $\theta$, the path integral generates an ordinary angular evolution. 
Denoting the $\theta$ associated Hamiltonian by $H_A$, one has the density matrix
\begin{equation}
 \widetilde\rho_A=e^{-2\pi H_A},
\label{eq:transfer}
\end{equation}
where
\begin{equation}
 H_A=\int_{Q_H}dV_H\,{\cal H}(x).
\label{eq:HA}
\end{equation}
Note that the CFT on the OST is the original planar CFT formulated on an open flat domain, with the same local fields and interactions. The closing surface is excluded from the manifold, and no physical boundary, boundary conditions, or additional boundary degrees of freedom are introduced. The finite parent state is defined directly by the Euclidean path integral on the OST, cut along a constant-$\theta$ slice. Since angular translations preserve the OST and the theory is $\theta$-independent, the usual transfer matrix construction gives eq.~(12). This state is not assumed to be the vacuum reduced state of an ordinary finite ball in planar coordinates. The conventional ball modular Hamiltonian is recovered in the singular limit as we are going to show below.
The modular Hamiltonian $K_A$ is defined by the normalized density matrix,
\begin{equation}
 \rho_A \equiv e^{-K_A} =\frac{e^{-2\pi H_A}}{Z_A},
 \qquad
 Z_A={\rm Tr}\,e^{-2\pi H_A},
  \qquad
 K_A=2\pi H_A+\log Z_A.
\label{eq:normalization}
\end{equation}
The finite parent therefore solves the operator part of the stationary modular flow problem directly. 
The trace $Z_A$   in general requires spectral, thermodynamic or response data. 
Note by using the replica trick, as shown in Ref.~\cite{Jiang:2026all}, 
the computation of entanglement entropy does not really need to calculate $Z_A$ itself. 

For the vacuum, continue to Lorentzian Rindler coordinates,
\begin{equation}
 t=\rho\sinh\eta,\qquad y=\rho\cosh\eta.
\label{eq:rindler}
\end{equation}
The angular generator becomes the boost generator. On $t=0$,
\begin{equation}
 H_R=\int_A d^{d-1}x\,\rho\,T_{00}
 =\int_{Q_H}dV_H\,T_{\eta\eta},
\label{eq:HR}
\end{equation}
so that the operator part of the modular Hamiltonian is $K_{A,\,{\rm OST}}^{\rm op}=2\pi H_R$. This is the finite-parent realization of the Bisognano--Wichmann weight.

\section{Finite Cavity and the CHM Singular Limit}
\label{sec:chm}

We saw that in the Fig. \ref{fig:ost} OST, the modular Hamiltonian eq. (\ref{eq:HR}) is  straightforward.
However, traditionally the modular Hamiltonians are  addressed not in the OST frame but 
in the singular limit $\epsilon\to 0$ of the Fig. \ref{fig:cavity} cavity frame which obscures the simplicity of the discussion. 
In order to have a clear comparison with the traditional approach, we map the Fig. \ref{fig:ost} OST to the Fig. \ref{fig:cavity} cavity frame.

\subsection{Exact finite conformal map}

\begin{figure}[t]
\centering
\includegraphics[width=0.3\textwidth]{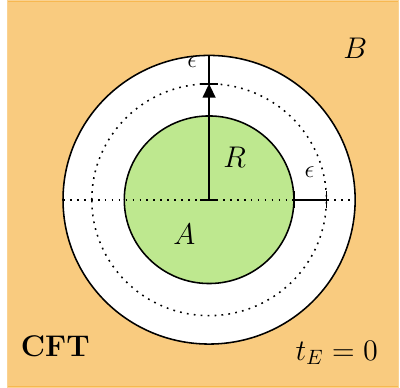}
\caption{The cavity representative on the $t_E=0$ slice. At finite $\epsilon$ the inner region $A$ and the exterior region $B$ are separated. The standard ball is reached only when the finite separation is collapsed.}
\label{fig:cavity}
\end{figure}

The finite hyperbolic cylinder (\ref{eq:globalH}) is conformally mapped to a spherical cavity in Fig. \ref{fig:cavity} by
\begin{equation}
 t_E=a\frac{\sin\theta}{\cosh u+\cos\theta},
 \qquad
 r=a\frac{\sinh u}{\cosh u+\cos\theta},
\label{eq:finitemap}
\end{equation}
with Weyl factor
\begin{equation}
 \Omega=\frac{a}{\cosh u+\cos\theta}.
\label{eq:Omega}
\end{equation}
The parameters satisfy the exact finite relations
\begin{equation}
 R=a\coth u_0,
 \qquad
 \epsilon=a\,\mathrm{csch}u_0,
 \qquad
 a=\sqrt{R^2-\epsilon^2},
 \qquad
 \cosh u_0=\frac{R}{\epsilon}.
\label{eq:finiteRelations}
\end{equation}
Note $0<\epsilon<R$ is a finite physical separation parameter. No small-$\epsilon$ expansion is used in the finite construction. A constant-$u$ surface obeys
\begin{equation}
 t_E^2+(r-a\coth u)^2=a^2\,\mathrm{csch}^2u.
\label{eq:constantusurface}
\end{equation}
For the outer edge $u=u_0$, write this sphere as
\begin{equation}
 t_E^2+(r-R)^2=\epsilon^2.
\label{eq:cavitysphere}
\end{equation}

The two intersections of the cavity sphere are $r=R\mp\epsilon$. The region corresponding to the finite hyperbolic ball is therefore bounded by $r<R-\epsilon$ on the interior branch. The familiar entangling sphere $r=R$ is not part of the finite configuration.

\subsection{Finite modular vector and modular Hamiltonian}

The angular vector $\partial_\theta$ pushes forward under eq.~\eqref{eq:finitemap} to
\begin{equation}
 \xi_E
 =\frac{a^2+t_E^2-r^2}{2a}\,\partial_{t_E}
 +\frac{t_Er}{a}\,\partial_r.
\label{eq:xiE}
\end{equation}
After Lorentzian continuation,
\begin{equation}
 \xi_a
 =\frac{a^2-t^2-r^2}{2a}\,\partial_t
 -\frac{tr}{a}\,\partial_r.
\label{eq:xiL}
\end{equation}
On $t=0$,
\begin{equation}
 \xi_a=\frac{a^2-r^2}{2a}\,\partial_t.
\label{eq:xislice}
\end{equation}
The operator part of the finite cavity vacuum modular Hamiltonian is therefore
\begin{equation}
 K_{A,{\rm cav}}^{\rm op}
 =2\pi\int_{r<R-\epsilon}d^{d-1}x\,
 \frac{a^2-r^2}{2a}T_{00}(0,\bm x).
\label{eq:Kcavity}
\end{equation}
Everything in eq.~\eqref{eq:Kcavity} is finite. The conventional ball has not yet been formed.
Note that the subscript ``cav''  indicates this is the modular Hamiltonian in the cavity frame, 
in comparison with the simple form eq. (\ref{eq:HR}) in the OST frame.

\subsection{CHM results from the singular endpoint}

Now take
\begin{equation}
 \epsilon\rightarrow0,
 \qquad a\rightarrow R,
 \qquad u_0\rightarrow\infty.
\label{eq:singularlimit}
\end{equation}
The finite spatial section $Q_H$ becomes the complete hyperbolic space $H^{d-1}$. Eq.~\eqref{eq:finitemap} becomes the standard CHM \cite{Casini:2011kv} map after the Lorentzian continuation $\theta=i\tau/R$,
\begin{equation}
 t=R\frac{\sinh(\tau/R)}{\cosh u+\cosh(\tau/R)},
 \qquad
 r=R\frac{\sinh u}{\cosh u+\cosh(\tau/R)}.
\label{eq:CHMmap}
\end{equation}
Removing the Weyl factor gives the hyperbolic cylinder
\begin{equation}
 ds_H^2=-d\tau^2+R^2\left(du^2+\sinh^2u\,d\Omega_{d-2}^2\right).
\label{eq:CHMcylinder}
\end{equation}
Since the parent Euclidean angle has period $2\pi$, the Euclidean hyperbolic time has period $2\pi R$. Thus the vacuum reduced state on the ball is unitarily equivalent to a thermal state on $\mathbb R\times H^{d-1}$ at
\begin{equation}
 T_0=\frac{1}{2\pi R}.
\label{eq:CHMtemperature}
\end{equation}
If $U$ denotes the conformal map between the two Hilbert-space representations,
\begin{equation}
 \rho_{B_R}
 =U^{-1}\frac{e^{-2\pi R H_H}}{Z_H}U.
\label{eq:CHMdensity}
\end{equation}
The modular flow becomes ordinary hyperbolic time translation, $\tau\rightarrow\tau+2\pi R s$, and the periodicity of the parent angular circle gives the corresponding thermal/KMS periodicity directly.

The same limit of eq.~\eqref{eq:xiL} gives the CHM \cite{Casini:2011kv} conformal Killing vector
\begin{equation}
 \xi_R
 =\frac{R^2-t^2-r^2}{2R}\,\partial_t
 -\frac{tr}{R}\,\partial_r,
\label{eq:CHMvector}
\end{equation}
and eq.~\eqref{eq:Kcavity} becomes
\begin{equation}
 K_{B_R}^{\rm op}
 =2\pi\int_{r<R}d^{d-1}x\,
 \frac{R^2-r^2}{2R}T_{00}(0,\bm x),
\label{eq:CHMK}
\end{equation}
which is the standard local spherical modular Hamiltonian in \cite{Casini:2011kv}.


Finally, under the limit  eq. (\ref{eq:singularlimit}),
the finite relation in eq.~\eqref{eq:finiteRelations} gives
\begin{equation}
 u_0=\operatorname{arccosh}\frac{R}{\epsilon}
 =\log\frac{2R}{\epsilon}+O(\epsilon^2/R^2).
\label{eq:UVIR}
\end{equation}
Thus the nontrivial CHM relation  between a short-distance cutoff near the sphere and a large-distance cutoff in hyperbolic space given in \cite{Casini:2011kv} is simply the asymptotic form of an exact finite geometric identity. At finite separation, $u_0$ is a physical size of the parent and $\epsilon$ is not a regulator.

\section{Analytic Continuation of Modular Flow and Stationary Sources}
\label{sec:sources}

\subsection{The center of the ball and its finite completion}

The finite parent makes a useful distinction between a genuine modular trajectory and a singular coordinate representation of that trajectory. This is visible already at the center of the cavity in Fig. \ref{fig:cavity}. Setting $u=0$ in eq.~\eqref{eq:finitemap} gives
\begin{equation}
 r=0,
 \qquad
 t_E=a\frac{\sin\theta}{1+\cos\theta}
 =a\tan\frac{\theta}{2}.
\label{eq:centerorbit}
\end{equation}
In the  Fig. \ref{fig:cavity} cavity coordinates the point $\theta=\pi$ is therefore sent to $t_E=\infty$. A single cavity or ball chart does not describe the full complexified orbit smoothly. On the Fig. \ref{fig:ost} OST, however, $u=0$ is an ordinary point of $Q_H$ and $\theta$ remains a smooth periodic coordinate. The complete orbit is present before the conformal map.

This is precisely the type of issue encountered in Ref.~\cite{Arias:2020mhe}. In their ball construction the analytical extension of the modular flow is ill-defined at the center $x^i=0$, and the complexified geometry is therefore written using $V_0=V\setminus\{0\}$. The finite-parent description shows that no physical degree of freedom needs to be deleted in order to define the modular evolution. The parent orbit is regular; the obstruction appears when that orbit is represented in a single ball coordinate patch. Thus the OST provides a global completion of the complexified modular flow. If one insists on working only in the conventional ball chart, the coordinate obstruction is of course still visible. The advantage of the finite parent is that this obstruction is no longer part of the definition of the state or of its modular evolution.

\subsection{Stationary Euclidean sources}

Refer to Fig. \ref{fig:ost},
consider a scalar primary ${\cal O}$ of dimension $\Delta$ and a local source $J_W$, 
\begin{equation}
 {\cal H}_J={\cal H}_0+J_W(x){\cal O}_W(x),
 \qquad \partial_\theta J_W=0.
\label{eq:sourceH}
\end{equation}
It is again straightforward to write down the modular flow in the OST frame
\begin{equation}
 H_J=H_R+\int_{Q_H}dV_H\,J_W{\cal O}_W,
 \qquad
 \widetilde\rho_J=e^{-2\pi H_J}.
\label{eq:sourceTransfer}
\end{equation}
The source may have arbitrary spatial dependence and need not be perturbatively small.
The conformal map acts on the source as
\begin{equation}
 J_{\rm cav}=\Omega^{\Delta-d}J_W.
\label{eq:sourceWeight}
\end{equation}
On the $t=0$ slice, the operator part of the finite cavity modular Hamiltonian is obtained by the  conformal map eq. (\ref{eq:finitemap}),
\begin{equation}
 K_{J,{\rm cav}}^{\rm op}
 =2\pi\int_{r<R-\epsilon}d^{d-1}x\,
 \frac{a^2-r^2}{2a}
 \left[T_{00}+J_{\rm cav}{\cal O}_{\rm cav}\right]_{t=0}.
\label{eq:sourceCavityK}
\end{equation}
The source is generally time dependent in ordinary Fig. \ref{fig:cavity} cavity coordinates; what remains exact is its stationarity with respect to the pushed-forward modular vector.

When the source has a regular ball limit, eq.~\eqref{eq:sourceCavityK} gives
\begin{equation}
 K_{J,B_R}^{\rm op}
 =2\pi\int_{r<R}d^{d-1}x\,
 \frac{R^2-r^2}{2R}
 \left[T_{00}+J_{B_R}{\cal O}\right]_{t=0},
\label{eq:sourceBallK}
\end{equation}
which is the stationary-source ball result discussed in Ref.~\cite{Arias:2020mhe}. 
The same local operator is therefore obtained in two ways: directly in the conventional singular representation, or as the endpoint of a nonsingular finite parent. The latter route makes the global angular evolution and the center completion manifest from the outset.

\subsection{Operator part and normalization}

The finite parent determines the modular generator in eqs.~\eqref{eq:HA}, \eqref{eq:Kcavity} and \eqref{eq:sourceCavityK}. It does not in general determine the normalization
\begin{equation}
 Z_A={\rm Tr}\,e^{-2\pi H_A}.
\label{eq:Zgeneral}
\end{equation}
This point is important. The earlier OST replica calculation of Ref.~\cite{Jiang:2026all} evaluates the von Neumann entropy from the response of the replicated geometry near $n=1$ which is crucial to circumvent the direct calculation of $Z_A$.
A direct fixed-Hamiltonian trace, by contrast, contains the full modular spectrum. 
Whenever the spectrum or an equation of state is independently known, however, the normalization can be computed in the ordinary way. The following examples illustrate this situation.

\section{Conserved Charge and a Four-Dimensional Dirac Benchmark}
\label{sec:charged}

\subsection{Finite-parent charged modular Hamiltonian}

Let the CFT contain a conserved global $U(1)$ charge $Q$ satisfying
\begin{equation}
 [H_R,Q]=0.
\label{eq:chargecommute}
\end{equation}
A constant angular chemical potential gives
\begin{equation}
 H_\mu=H_R-\mu Q,
\label{eq:Hmu}
\end{equation}
and hence
\begin{equation}
 \rho_\mu=\frac{e^{-2\pi(H_R-\mu Q)}}{Z_\mu},
 \qquad
 K_\mu=2\pi(H_R-\mu Q)+\log Z_\mu.
\label{eq:chargedrho}
\end{equation}
The charge term generates an internal $U(1)$ rotation in addition to the spacetime modular flow. The operator is local but not purely geometric in the strict stress-tensor sense.

At fixed physical period,
\begin{equation}
 \partial_\mu\log Z_\mu=2\pi\langle Q\rangle_\mu,
 \label{eq:U1muRoute}
\end{equation}
which determines a normalization \emph{ratio}
\begin{equation}
 \log\frac{Z_\mu}{Z_0}
 =2\pi\int_0^\mu d\mu'\,\langle Q\rangle_{\mu'}.
 \label{eq:U1logZratio}
\end{equation}
This relation does not require the absolute uncharged normalization.  Correspondingly, the entropy difference is
\begin{align}
 S_\mu-S_0
 ={}&2\pi\left(\langle H_R\rangle_\mu-\langle H_R\rangle_0
 -\mu\langle Q\rangle_\mu\right)
 +\log\frac{Z_\mu}{Z_0}.
 \label{eq:generalDeltaSmu}
\end{align}
Thus charged entropy differences are available whenever the ordinary charge and energy data are independently known.  Charged R\'enyi and symmetry-resolved quantities for spherical regions have long been formulated through background Wilson lines and the hyperbolic-cylinder map \cite{Belin:2013uta,Dowker:2015dpd}; here they serve as a benchmark for the finite-parent organization.

\subsection{Complete-hyperbolic Dirac benchmark}\label{subsec:Dirac4}

For a quantitative check we now take the singular complete-hyperbolic limit $u_0\to\infty$, where the standard spectrum is known.  Consider a four-component massless Dirac field in $d=4$ on $S^1_\beta\times H^3$.  The Plancherel measure per unit hyperbolic volume is \cite{Camporesi:1994ga,Belin:2013uta}
\begin{equation}
 \frac{dN}{V_{H^3}}
 =\frac{d\lambda}{\pi^2}\left(\lambda^2+\frac14\right),
 \qquad \lambda\ge0.
 \label{eq:DiracDensity}
\end{equation}
After subtracting the temperature-independent zero-point term,
\begin{align}
 \frac{\log Z(\beta,\mu)}{V_{H^3}}
 =&\int_0^\infty\frac{d\lambda}{\pi^2}
 \left(\lambda^2+\frac14\right)
 \nonumber\\
 &\times\left[\log\left(1+e^{-\beta(\lambda-\mu)}\right)
 +\log\left(1+e^{-\beta(\lambda+\mu)}\right)\right].
 \label{eq:DiracLogZIntegral}
\end{align}
The integral gives
\begin{equation}
 \frac{\log Z(\beta,\mu)}{V_{H^3}}
 =\frac{7\pi^2}{180\beta^3}
 +\frac{1}{24\beta}
 +\frac{\mu^2}{6\beta}
 +\frac{\beta\mu^2}{8\pi^2}
 +\frac{\beta\mu^4}{12\pi^2}
 +C\beta,
 \label{eq:DiracLogZClosed}
\end{equation}
where $C$ is a vacuum-energy convention.  At $\beta=2\pi$,
\begin{equation}
 q_\mu\equiv\frac{\langle Q\rangle_\mu}{V_{H^3}}
 =\frac{\mu+\mu^3}{3\pi^2},
 \label{eq:Diracq}
\end{equation}
and
\begin{equation}
 \frac{\langle H_R\rangle_\mu-\langle H_R\rangle_0}{V_{H^3}}
 =\frac{\mu^2+\mu^4}{4\pi^2}.
 \label{eq:Diraceps}
\end{equation}
Eq.~\eqref{eq:U1logZratio} gives
\begin{equation}
 \frac{\log Z_\mu-\log Z_0}{V_{H^3}}
 =\frac{\mu^2}{3\pi}+\frac{\mu^4}{6\pi},
 \label{eq:DiracDeltaLogZ}
\end{equation}
and therefore the entropy difference is
\begin{equation} 
 \frac{S_\mu-S_0}{V_{H^3}}=\frac{\mu^2}{6\pi}.
 \label{eq:DiracDeltaS}
\end{equation}
Using the known uncharged result \cite{Belin:2013uta},
\begin{equation}
 \frac{S_0}{V_{H^3}}=\frac{11}{180\pi},
 \label{eq:DiracS0}
\end{equation}
one recovers
\begin{equation}
 \frac{S_\mu}{V_{H^3}}
 =\frac{11}{180\pi}+\frac{\mu^2}{6\pi}.
 \label{eq:DiracSmuAbsolute}
\end{equation}
This agrees with the $n\to1$ charged free-Dirac result of Ref.~\cite{Belin:2013uta} after the standard analytic continuation between Euclidean and thermodynamic chemical potentials. Some computational  details are collected in Appendix~\ref{app:dirac}.

\subsection{Finite cavity and singular charged ball}
At finite parent size the charge term is already present in eq.~(\ref{eq:chargedrho}).
Mapping to the cavity configuration by eq. (\ref{eq:finitemap}), we have the operator sector of modular Hamiltonian in the cavity coordinate
\begin{equation}
 K_{\mu,{\rm cav}}^{\rm op}
 =2\pi\int_{r<R-\epsilon}d^{d-1}x\,
 \frac{a^2-r^2}{2a}T_{00}
 -2\pi\mu Q_A.
\label{eq:chargedcavity}
\end{equation}
In the singular limit,
\begin{equation}
 K_{\mu,B_R}^{\rm op}
 =2\pi\int_{r<R}d^{d-1}x\,
 \frac{R^2-r^2}{2R}T_{00}
 -2\pi\mu Q_A.
\label{eq:chargedball}
\end{equation}
This is the operator structure underlying the charged spherical constructions of Ref.~\cite{Belin:2013uta}. The finite OST supplies its nonsingular ancestor.

\section{An Inhomogeneous Scalar Shell}
\label{sec:shell}

The advantage of the OST finite parent construction is not restricted to homogeneous states. 
In this section, we provide a $\theta$ translation invariant but spatial inhomogeneous example, namely a scalar shell.
Consider the conformal scalar in $d=3$ on
\begin{equation}
 ds_W^2=d\theta^2+du^2+\sinh^2u\,d\varphi^2,
 \qquad {\cal R}=-2,
\label{eq:shellmetric}
\end{equation}
with conformal coupling $\xi=1/8$. Add the stationary quadratic source
\begin{equation}
 \delta I_g=\frac12\int d\theta\,dV_H\,g(u):\!\Phi^2\!:
\label{eq:shellsource}
\end{equation}
with
\begin{equation}
 g(u)=g_0\,\Theta(u-u_1)\Theta(u_2-u),
 \qquad 0<u_1<u_2<u_0.
\label{eq:shellprofile}
\end{equation}
The OST finite parent modular Hamiltonian is immediately
\begin{equation}
 H_g=\frac12\int dV_H\left[\Pi^2+\Phi{\cal K}_g\Phi\right],
 \qquad
 {\cal K}_g=-\nabla_{H^2}^2-\frac14+g(u).
\label{eq:shellH}
\end{equation}
This result is exact for arbitrary $g_0$. No perturbative source expansion is needed to identify the operator.

At finite $u_0$ the important result is eq.~\eqref{eq:shellH}. 
To evaluate the normalization
trace, i.e $Z={\rm Tr}\, e^{-2\pi H_g}$, one must solve the corresponding spectral problem. In the singular complete-hyperbolic limit $u_0\rightarrow\infty$, the localized shell becomes an ordinary scattering problem on $H^2$. If
\begin{equation}
 {\cal K}_g f_\alpha=\omega_\alpha^2 f_\alpha,
\end{equation}
the bosonic entropy functional is
\begin{equation}
 S_g={\rm Tr}\,s(2\pi\sqrt{{\cal K}_g}),
 \qquad
 s(x)=\frac{x}{e^x-1}-\log(1-e^{-x}).
\label{eq:shellentropyfunctional}
\end{equation}
For a compact shell the  entropy difference can be expressed through partial wave phase shifts,
\begin{equation}
 \Delta S_{\rm shell}
 =\sum_{m\in\mathbb Z}\int_0^\infty\frac{dp}{\pi}
 \frac{d\delta_m(p)}{dp}\,s(2\pi p).
\label{eq:shellphase}
\end{equation}
For $g_0\ge0$, the shell raises the oscillator frequencies and therefore
\begin{equation}
 \frac{dS_g}{dg_0}\le0.
\label{eq:shellmonotonic}
\end{equation}
Some computational  details are given in Appendix \ref{app:shell}.

The phase-shift method itself is standard; the point is that the finite parent has already supplied the local modular Hamiltonian before the scattering calculation is performed. A Gaussian covariance-matrix method could of course reconstruct the same entropy from two-point data \cite{Srednicki:1993im,Casini:2009sr}. The two methods organize the problem differently.

For the operator $:\!\Phi^2\!:$, $\Delta=1$ in the three-dimensional free scalar CFT. Under the finite conformal map,
\begin{equation}
 g_{\rm cav}(t_E,r)=\Omega^{-2}(t_E,r)g(u(t_E,r)).
\label{eq:shellmap}
\end{equation}
A fixed shell surface $u=u_i$ becomes
\begin{equation}
 t_E^2+\left(r-a\coth u_i\right)^2=a^2\,\mathrm{csch}^2u_i.
\label{eq:shellcircle}
\end{equation}
Thus the inhomogeneous state has a completely finite cavity realization. The source may be time dependent in ordinary cavity coordinates but remains stationary under angular rotation. The conventional ball state is obtained only after the finite density matrix has been constructed and the final singular limit is taken.

\section{Summary and Discussion}
\label{sec:discussion}

We have studied modular flow from the finite-parent point of view. The basic observation is geometric. On an open solid torus the angular action is free and the angular path integral is an ordinary transfer evolution. After a Weyl transformation the spatial section is a finite hyperbolic ball. The resulting modular generator is therefore defined before an entangling fixed surface, a hyperbolic infinity, or a UV cutoff is introduced. The advantage of the open solid torus is that the modular Hamiltonian along the angular $\theta$ is very easy to understand. As long as the system has a $U(1)$ invariance, a local modular Hamiltonian can be straightforwardly defined. Then an exact conformal map can transports this finite construction to a spherical cavity. After taking the singular limits, one achieves the conventional planar CFT results.

\vspace*{3.0ex}
\begin{acknowledgments}
\paragraph*{Acknowledgments.}
This work is supported by NSFC (Grant No. 12275184, 12675088).
\end{acknowledgments}

\bibliographystyle{unsrturl}
\bibliography{Finite_Modular_refs}

\appendix

\section{Hyperbolic Coordinates of the Finite Parent}
\label{app:coordinates}

The upper-half-space model of unit $H^{d-1}$ is
\begin{equation}
 ds_H^2=\frac{d\rho^2+d\bm{x}_\perp^{\,2}}{\rho^2},
 \qquad \rho>0.
\label{eq:appPoincare}
\end{equation}
Choose the hyperbolic center $P_c=(\rho_c,\bm0)$. The geodesic distance satisfies Eq.~\eqref{eq:hyperdistance}. Introducing an angle $\alpha$ for the transverse direction, we obtain
\begin{equation}
 \rho=\frac{\rho_c}{\cosh u-\sinh u\cos\alpha},
\label{eq:appRho}
\end{equation}
\begin{equation}
 |\bm{x}_\perp|
 =\frac{\rho_c\sinh u\sin\alpha}
 {\cosh u-\sinh u\cos\alpha}.
\label{eq:appXperp}
\end{equation}
Substitution into Eq.~\eqref{eq:appPoincare} yields
\begin{equation}
 ds_H^2=du^2+\sinh^2u\,d\Omega_{d-2}^2.
\end{equation}
For the Euclidean sphere in Eq.~\eqref{eq:ostdomain},
\begin{equation}
 R_c=\frac{R_1+R_2}{2},\qquad
 R_0=\frac{R_2-R_1}{2}.
\end{equation}
A geodesic sphere centered at $\rho_c$ obeys
\begin{equation}
 R_c=\rho_c\cosh u_0,
 \qquad
 R_0=\rho_c\sinh u_0,
\end{equation}
which gives $\rho_c=\sqrt{R_1R_2}$ and Eq.~\eqref{eq:u0}.

\section{Finite Cavity Map and Source Weight}
\label{app:cavity}

Let $D_0=\cosh u+\cos\theta$. From Eq.~\eqref{eq:finitemap},
\begin{equation}
 \frac{\partial t_E}{\partial\theta}
 =a\frac{1+\cosh u\cos\theta}{D_0^2},
 \qquad
 \frac{\partial r}{\partial\theta}
 =a\frac{\sinh u\sin\theta}{D_0^2}.
\end{equation}
Eliminating $(u,\theta)$ gives
\begin{equation}
 \frac{\partial t_E}{\partial\theta}
 =\frac{a^2+t_E^2-r^2}{2a},
 \qquad
 \frac{\partial r}{\partial\theta}=\frac{t_Er}{a},
\end{equation}
which proves Eq.~\eqref{eq:xiE}. The constant-$u$ relation Eq.~\eqref{eq:constantusurface} follows directly from the same map.

For a scalar primary ${\cal O}_\Delta$, conformal invariance of the source coupling gives
\begin{equation}
 \int d^d x\sqrt{g_W}\,J_W{\cal O}_W
 =\int d^d x\sqrt{g_{\rm cav}}\,J_{\rm cav}{\cal O}_{\rm cav},
\end{equation}
with $g_{\rm cav}=\Omega^2g_W$ and ${\cal O}_{\rm cav}=\Omega^{-\Delta}{\cal O}_W$. Hence Eq.~\eqref{eq:sourceWeight} follows.

\section{Four-dimensional Dirac thermodynamics on $S^1\times H^3$}
\label{app:dirac}

We collect details of the calculation in Sec.~\ref{subsec:Dirac4}. For a four-component massless Dirac field on unit $H^3$, the spectral density per unit hyperbolic volume is
\begin{equation}
 \rho(\lambda)=\frac{1}{\pi^2}\left(\lambda^2+\frac14\right).
 \label{eq:appDiracDensity}
\end{equation}
The finite-temperature part of the grand potential is therefore eq.~\eqref{eq:DiracLogZIntegral}. Set $x=\beta\lambda$ and $\alpha=\beta\mu$. The two Fermi integrals needed are
\begin{equation}
 \int_0^\infty dx\,
 \left[\log(1+e^{-x+\alpha})+\log(1+e^{-x-\alpha})\right]
 =\frac{\pi^2}{6}+\frac{\alpha^2}{2},
 \label{eq:Fermi0}
\end{equation}
\begin{align}
 \int_0^\infty dx\,x^2
 \left[\log(1+e^{-x+\alpha})+\log(1+e^{-x-\alpha})\right]
 =&\ \frac{7\pi^4}{180}
 +\frac{\pi^2\alpha^2}{6}
 +\frac{\alpha^4}{12}.
 \label{eq:Fermi2}
\end{align}
They give
\begin{equation}
 \frac{\log Z}{V_H}
 =\frac{7\pi^2}{180\beta^3}
 +\frac{1}{24\beta}
 +\frac{\mu^2}{6\beta}
 +\frac{\beta\mu^2}{8\pi^2}
 +\frac{\beta\mu^4}{12\pi^2}
 +C\beta.
 \label{eq:appDiracLogZ}
\end{equation}
The constant $C$ represents the scheme-dependent linear vacuum contribution. It drops from
\begin{equation}
 q(\beta,\mu)
 =\frac{1}{\beta V_H}\partial_\mu\log Z
 =\frac{\mu}{3\beta^2}
 +\frac{\mu}{4\pi^2}
 +\frac{\mu^3}{3\pi^2},
 \label{eq:appDiracqBeta}
\end{equation}
and from entropy differences. At $\beta=2\pi$ this becomes eq.~\eqref{eq:Diracq}.

The grand modular-energy density is
\begin{align}
 u(\beta,\mu)
 =-\frac1{V_H}\partial_\beta\log Z
 =&\ \frac{7\pi^2}{60\beta^4}
 +\frac{1}{24\beta^2}
 +\frac{\mu^2}{6\beta^2}
 \nonumber\\
 &-\frac{\mu^2}{8\pi^2}
 -\frac{\mu^4}{12\pi^2}-C.
 \label{eq:appDiracuBeta}
\end{align}
The ordinary energy is $\varepsilon=u+\mu q$. Taking the difference from $\mu=0$ at $\beta=2\pi$ yields eq.~\eqref{eq:Diraceps}. Finally,
\begin{equation}
 \frac{S(\beta,\mu)}{V_H}
 =\beta u(\beta,\mu)+\frac{\log Z(\beta,\mu)}{V_H}
 \label{eq:appDiracEntropyDef}
\end{equation}
contains no $C$. At $\beta=2\pi$,
\begin{equation}
 \frac{S(2\pi,\mu)}{V_H}
 =\frac{11}{180\pi}+\frac{\mu^2}{6\pi},
 \label{eq:appDiracEntropy}
\end{equation}
which is Eq.~\eqref{eq:DiracSmuAbsolute}.

\section{Quadratic Shell and Spectral Shift}
\label{app:shell}

Expand
\begin{equation}
 \Phi(\theta,u,\varphi)
 =\sum_{m\in\mathbb Z}e^{im\varphi}\Phi_m(\theta,u).
\end{equation}
The radial equation in the complete-hyperbolic limit is
\begin{equation}
 \left[-\partial_u^2-\coth u\,\partial_u
 +\frac{m^2}{\sinh^2u}-\frac14+g(u)\right]\psi_m
 =p^2\psi_m.
\label{eq:appShellRadial}
\end{equation}
For the square shell, solutions in the three radial regions are matched by continuity of $\psi_m$ and its radial derivative at $u_1,u_2$. The asymptotic solution defines the phase shift $\delta_m(p)$. The relative density of states is
\begin{equation}
 \Delta\rho_m(p)=\frac{1}{\pi}\frac{d\delta_m(p)}{dp},
\end{equation}
which gives eq.~\eqref{eq:shellphase}. This is the usual spectral-shift mechanism underlying scattering formulations of thermal partition functions \cite{Dashen:1969ep}.

Monotonicity follows from the Hellmann--Feynman theorem. For a normalized mode,
\begin{equation}
 \frac{d\omega_\alpha}{dg_0}
 =\frac{1}{2\omega_\alpha}
 \int_{u_1<u<u_2}dV_H\,|f_\alpha|^2\ge0,
\end{equation}
while
\begin{equation}
 s'(x)=-\frac{x e^x}{(e^x-1)^2}<0.
\end{equation}
Therefore $dS_g/dg_0\le0$.

\end{document}